# Universal Properties of 2-Port Scattering, Impedance and Admittance Matrices of Wave Chaotic Systems.

Sameer Hemmady [1,2,3,4], Xing Zheng [5], James Hart [1], Thomas M. Antonsen Jr. [1,2,3], Edward Ott [1,2,3] and Steven M. Anlage [1,2,4].
[1]Department of Physics, University of Maryland, College Park, MD 20742-4111, U.S.A.

**[Dated: July 10$^{th}$, 2006].**

**Abstract:**
Statistical fluctuations in the eigenvalues of the scattering, impedance and admittance matrices of 2-Port wave-chaotic systems are studied experimentally using a chaotic microwave cavity. These fluctuations are universal in that their properties are dependent only upon the degree of loss in the cavity. We remove the direct processes introduced by the non-ideally coupled driving ports through a matrix-normalization process that involves the radiation-impedance matrix of the two driving ports. We find good agreement between the experimentally obtained marginal probability density functions (PDFs) of the eigenvalues of the normalized impedance, admittance and scattering matrix and those from Random Matrix Theory (RMT). We also experimentally study the evolution of the joint PDF of the eigenphases of the normalized scattering matrix as a function of loss. Experimental agreement with the theory by Brouwer and Beenakker for the joint PDF of the magnitude of the eigenvalues of the normalized scattering matrix is also shown.

**PACS Number (s) : 05.45.Mt, 03.65.Nk,11.55.-m,03.50.De, 04.30.Nk**

## I A: Introduction:

The scattering of short-wavelength waves inside enclosures manifests itself in several fields of physics and engineering such as quantum dots [1], atomic nuclei [2], acoustic resonators [3, 4], electromagnetic compatibility [5], etc. Of particular interest is the case when the ray trajectories within the enclosure show chaotic dynamics in the classical limit. This interest has spawned the field of "wave chaos" (or "quantum chaos"), and has attracted much theoretical and experimental work [6, 7] to understand its nature. On account of the small wavelength of the scattered waves, as compared to the characteristic length-scale of the enclosure, the response of these systems exhibit extreme sensitivity to small changes in configuration, driving frequency, nature of driving ports, ambient conditions such as temperature, etc. Thus, an intimate knowledge of the response of any such system for a given well-defined stimulus or system configuration will not provide any foresight in predicting the response of a similar system when the stimulus or system configuration is slightly altered. This calls for a statistical approach to quantify the nature of such wave-chaotic systems.

In this regard, Random Matrix Theory (RMT) [8] has proved to be an integral tool in predicting universal statistical aspects of wave chaotic systems. It has been conjectured that in the short-wavelength regime, RMT can be used to model wave-chaotic systems. In particular, the statistics of systems that show Time-Reversal Symmetry are conjectured to be described by the Gaussian Orthogonal Ensemble (GOE) of random matrices, while the statistics of systems showing Broken Time-Reversal Symmetry are conjectured to be described by the Gaussian Unitary Ensemble (GUE) of random matrices. There is also a third random matrix ensemble corresponding to certain systems with spin-interactions (Gaussian Symplectic Ensemble). RMT provides a potential framework for uncovering universal statistical properties of short-wavelength wave chaotic systems (e.g. Ericson fluctuations in nuclear scattering [7, 9] and universal conductance fluctuations (UCF) in quantum-transport systems [10] ).

Since the applicability of RMT and the concomitant universal statistics is conjectural rather than rigorous, and since this conjectured applicability is said to be asymptotic in the limit of wavelength small compared to the system size, it is important to test the RMT conjecture against results obtained for specific real situations.

Experimentally, however, validating the applicability of RMT has always proved challenging. One of the most common problems encountered by experimentalists is the presence of non-universal, system-specific artifacts introduced into the measured data by the experimental apparatus. These are generally referred to as the "direct processes", as opposed to the "equilibrated processes" which describe the chaotic scattering within the system [11]. A typical example presents itself while measuring the statistical fluctuations in the scattering of microwaves through cavities with chaotic ray dynamics. These fluctuations are studied by exciting the cavity through coupled ports and observing the response (reflection and transmission) for a given excitation. Generally, it is not possible to perfectly couple (ideally match) the ports to the cavity at all frequencies. Here by "perfect coupling" we refer to the situation in which there is no prompt reflection of a wave incident from an incoming channel on a cavity port. Thus, such a wave would be entirely transmitted into the cavity, and any reflection coefficient measured from that port is the result of waves that have entered the cavity, bounced around, and subsequently been reflected back toward the port. (Subsequently in this section and section I (b), we give a more precise definition of perfect coupling in terms of the port radiation impedance and the characteristic impedance of the incoming channel.) We refer to the deviation from perfect coupling as "mismatch". This mismatch, which is strongly determined by the geometry of the port, manifests itself as systematic fluctuations in the measured data. The result is that the measured data depends on the non-universal, direct processes of the ports, as well as the underlying universal, equilibrated processes of the chaotic scattering system.

Several approaches have been formulated to account for these direct processes [12, 13, 14] of which the "Poisson Kernel" approach introduced by Mello, Pereyra and Seligman is of special mention. Based on an information-theoretic model, the "Poisson Kernel" characterizes the direct processes between the ports and the cavity by the ensemble-averaged scattering matrix $<<\vec{S}>>$. In order to apply this theory to a specific real situation, it is thus necessary to obtain a quantity that plays the role of the ensemble average $<<\vec{S}>>$ appropriate to that specific system. For example, one scheme proposed for determining

2: Department of Electrical and Computer Engineering
3: Institute for Research in Electronics and Applied Physics.
4: Center for Superconductivity Research.
5: Now at George Washington University.

such a surrogate for $<<\vec{S}>>$ for a specific system used system configuration averaging. We denote this surrogate for $<<\vec{S}>>$ as $<\vec{S}>$. Averaging over configurations, however, may suffer from excessive statistical error if the number of configurations averaged over is insufficiently large. Thus, to improve the estimate of the scattering coefficient statistics, Refs. [15, 16], which treat one port (scalar $S$) scatterers, make use of an ergodic hypothesis [17, 18] to include an additional running average over frequency ranges that include many resonances but are sufficiently small that the scattering coefficient statistics can be assumed to be nearly constant (i.e., a frequency range where the port coupling strengths are nearly constant). Using this approach, Refs. [15, 16] have investigated the universal fluctuations in the reflection coefficient of 1-port wave-chaotic microwave cavities. This was shown to produce favorable results for 1-port systems when compared with RMT predictions. We note, however, that the analysis is highly dependent on the accuracy of the experimentally-obtained $<S>$, which is prone to statistical errors.

The situation can become even more complicated when dealing with $N$ ports. In the recent 2-port paper, Ref. [19], the authors circumvent such problems by taking careful steps to ensure that the driving ports are nearly perfectly-coupled to the cavity in the frequency range where the data is analyzed. In doing so, Ref. [19] achieves good agreement between the experimental results for the fluctuations in the transmission coefficient, and the RMT predictions for time-reversal-symmetric and for broken-time-reversal-symmetric cavities. We note, however, that Ref. [19] is for the case of perfectly coupled ports and that it is desirable to also deal with arbitrary port couplings.

In Ref. [20, 21] a novel method to characterize the direct processes between the cavity and the driving ports was introduced. This method, which is motivated by electromagnetic-wave propagation inside complex enclosures, makes use of impedances to characterize the direct-processes rather than the ensemble-averaged scattering matrix as in Ref. [12]. For an $N$-port scattering system, the Scattering Matrix $\vec{S}$ models the scattering region of interest in terms of an $NxN$ complex-valued matrix. Specifically, it expresses the amplitudes of the $N$ outgoing scattered waves ($\tilde{b}$) in terms of the $N$ incoming waves ($\tilde{a}$) at the location of each port (i.e., $\tilde{b} = \vec{S}\tilde{a}$). The impedance matrix $\vec{Z}$, on the other hand, is a quantity which relates the complex voltages ($\tilde{V}$) at the $N$ driving ports to the complex currents ($\tilde{I}$) in the $N$ ports (i.e. $\tilde{V} = \vec{Z}\tilde{I}$). The matrices $\vec{S}$ and $\vec{Z}$ are related through the bilinear transformation, $\vec{S} = \vec{Z}_o^{1/2}(\vec{Z}+\vec{Z}_o)^{-1}(\vec{Z}-\vec{Z}_o)\vec{Z}_o^{-1/2}$ where $\vec{Z}_o$ is the $NxN$ real, diagonal matrix whose elements are the characteristic impedances of the waveguide (or transmission line) input channels at the $N$ driving ports. Like $\vec{S}$, $\vec{Z}$ is also a well-established physical quantity in quantum mechanics. Just as the elements of $\vec{S}$ represent the transition probabilities from one state to the other in a quantum scattering system, $\vec{Z}$ is an electromagnetic analog to Wigner's Reaction Matrix [22], which linearly relates the wave function to its normal derivative at the boundary separating the scattering region from the outside world.

References [20, 21] have shown that the direct processes can be quantified by the "radiation impedance" of the driving-ports. For a cavity driven by a single port, the radiation impedance of the port is that impedance observed at the reference plane of the port which retains its coupling geometry but has the distant walls of the cavity moved out to infinity and an outward radiation condition imposed. Experimentally, this can be realized by lining the walls that are distant from the port with microwave absorber. The one-port radiation impedance, denoted $Z_{rad}$, is thus a frequency-dependant, complex scalar quantity which depends only on the local structure of the port and is not influenced by the shape of the distant cavity boundaries; $Z_{rad} = \text{Re}[Z_{rad}] + i\text{Im}[Z_{rad}]$, where $\text{Re}[Z_{rad}]$ is the "radiation resistance" which quantifies the energy dissipated in the far-field of the radiating port, and $\text{Im}[Z_{rad}]$ is the "radiation reactance" which arises from energy stored in the near-field of the radiating port. $Z_{rad}$ thus presents a non-statistical experimentally viable way to quantify the direct processes in a wave-chaotic system for any given port geometry; without resorting to averaging.

This "radiation impedance" approach has been used successfully by Ref. [23, 24] for a two-dimensional, wave-chaotic resonator which is driven by a single port. In Ref. [23], the authors used the measured radiation impedance $Z_{rad}$ of the driving port to normalize the measured cavity data $Z$ and recover the universal normalized cavity impedance $z = (Z - i\text{Im}[Z_{rad}])/\text{Re}[Z_{rad}]$. This normalized impedance $z$ represents the scalar cavity impedance when the driving port is perfectly coupled to the cavity (i.e., $Z_{rad} = Z_o$, where $Z_o$ is the characteristic impedance of the transmission line connected to the port). Reference [23] has shown that the Probability Density Functions (PDFs) of $\text{Re}[z]$ and $\text{Im}[z]$ are independent of the geometry of the coupling port; but rather depend solely on the degree of quantifiable loss in the system. Reference [24] carried forward the one-port results of Ref. [23] to relate $z$ to the normalized scattering coefficient $s = (z-1)/(z+1)$, which describes the scattering fluctuations in a cavity which is perfectly-coupled to its driving-port. This expression for $s$ follows from classical electromagnetic theory and relates the scattering coefficient at the plane of measurement with the load impedance [25] on a transmission line. In Refs. [26, 27] a similar expression for the normalized scattering coefficient is given in terms of Wigner's Reaction Matrix ($K$), i.e. $\breve{s} = (1-iK)/(1+iK)$ with $z = iK$. The two quantities $s$ and $\breve{s}$ differ in phase by $\pi$ radians. This extra phase contribution can be easily absorbed into the uniformly-distributed phase of $s$ (Ref.[24]) thereby yielding identical statistical descriptions for $s$ and $\breve{s}$.

Both Refs. **[23, 24]** have been able to experimentally verify several universal statistical properties of $z$ and $s$, which are in good agreement with numerical results from Random Matrix Theory. The "radiation impedance" approach to characterizing the direct processes in wave-chaotic systems has also been independently investigated by Ref. **[28]** in a 1-port, three-dimensional, mode-stirred chamber. References **[20, 21]** have also shown that in the limit that the number of samples determining the average $<S>$ goes to infinity, $S_{rad} = <S>$ thereby making contact with the Poisson Kernel approach, where $S_{rad} = (Z_{rad} - Z_o)/(Z_{rad} + Z_o)$. This connection has been experimentally established in Refs. **[24, 29]**.

## I B: Extending the "Radiation Impedance" approach to 2-Port Wave-Chaotic Systems.

Here, we experimentally extend the "radiation impedance" approach of Refs. **[20, 21]** to two-port chaotic cavities. In general, for a $N$-port system, the radiation impedance is now an $N \times N$ complex-valued, symmetric matrix ($\vec{\vec{Z}}_{rad}$). If the $N$ ports are very far apart, $\vec{\vec{Z}}_{rad}$ is diagonal, but we do not assume that here. Reference **[21]** has shown that the measured $NxN$ impedance matrix of a $N$-port, wave-chaotic cavity ($\vec{\vec{Z}}$) has a mean-part given by the radiation impedance matrix ($\vec{\vec{Z}}_{rad}$) and a universal fluctuating part ($\vec{\vec{z}}$), which is scaled by the radiation resistance matrix ($\text{Re}[\vec{\vec{Z}}_{rad}]$). Thus,

$$\vec{\vec{Z}} = i\,\text{Im}[\vec{\vec{Z}}_{rad}] + (\text{Re}[\vec{\vec{Z}}_{rad}])^{1/2}\vec{\vec{z}}(\text{Re}[\vec{\vec{Z}}_{rad}])^{1/2}. \quad (1)$$

From (1), we can easily extract $\vec{\vec{z}}$,

$$\vec{\vec{z}} = (\text{Re}[\vec{\vec{Z}}_{rad}])^{-1/2}(\vec{\vec{Z}} - i\,\text{Im}[\vec{\vec{Z}}_{rad}])(\text{Re}[\vec{\vec{Z}}_{rad}])^{-1/2}. \quad (2)$$

The normalized scattering matrix $\vec{\vec{s}}$ is,

$$\vec{\vec{s}} = (\vec{\vec{z}} - \vec{\vec{1}})(\vec{\vec{z}} + \vec{\vec{1}})^{-1}, \quad (3)$$

where $\vec{\vec{1}}$ is the $NxN$ identity matrix.

The normalized scattering matrix $\vec{\vec{s}}$ can also be obtained from the cavity scattering matrix $\vec{\vec{S}}$ and the radiation scattering matrix $\vec{\vec{S}}_{rad}$ by converting these quantities to the cavity and radiation impedances, $\vec{\vec{Z}}$ and $\vec{\vec{Z}}_{rad}$, respectively through

$$\vec{\vec{Z}} = \vec{\vec{Z}}_o^{1/2}(\vec{\vec{1}} + \vec{\vec{S}})(\vec{\vec{1}} - \vec{\vec{S}})^{-1}\vec{\vec{Z}}_o^{1/2} \text{ and}$$
$$\vec{\vec{Z}}_{rad} = \vec{\vec{Z}}_o^{1/2}(\vec{\vec{1}} + \vec{\vec{S}}_{rad})(\vec{\vec{1}} - \vec{\vec{S}}_{rad})^{-1}\vec{\vec{Z}}_o^{1/2}, \quad (4)$$

and by then using Eqs.(2) and (3). The matrix $\vec{\vec{Z}}_o$ is a real diagonal matrix whose elements are the characteristic impedances of the transmission lines connected to the driving ports.

The normalized quantities $\vec{\vec{z}}$ and $\vec{\vec{s}}$ represent the impedance and scattering matrix when the $N$ ports are perfectly coupled to the cavity, i.e., when $\vec{\vec{Z}}_{rad} = \vec{\vec{Z}}_o$. Since, in general, $\vec{\vec{Z}}_{rad}$ is a smoothly varying function of frequency and of the coupling-port structure, Eqs. (2) and (3) yield the perfectly-coupled (ideally matched) impedance and scattering matrix over any arbitrarily large range of frequency and for any port geometry.

Reference **[21]** predicts that the PDFs of the eigenvalues of $\vec{\vec{z}}$ which are contained in the diagonal matrix $\vec{\vec{\lambda}}_{\vec{z}}$, and PDFs of the eigenvalues of $\vec{\vec{s}}$ which are contained in the diagonal matrix $\vec{\vec{\lambda}}_{\vec{s}}$ are qualitatively similar to the PDFs of $z$ and $s$ in the 1-port case and that they are dependent only on the loss-parameter of the cavity. Loss is quantified by the expression $k^2/(\Delta k_n^2 Q)$ **[20, 21]**. Here, $k = 2\pi f/c$ is the wavenumber for the incoming frequency $f$ and $\Delta k_n^2$ is the mean-spacing of the adjacent eigenvalues of the Helmholtz operator, $\nabla^2 + k^2$, as predicted by the Weyl Formula **[30]** for the closed system. The use of the Weyl Formula here is conventionally accepted for lack of a more complete treatment which is applicable to open systems or to systems with high absorption (as discussed in this paper). The quantity Q represents the loaded quality factor of the cavity. The quantity $k^2/(\Delta k_n^2 Q)$ thus represents the ratio of the frequency width of the cavity resonances due to distributed losses to the average spacing between resonant frequencies. Reference **[20]** also predicts that the variance ($\sigma^2$) of the PDFs of $\text{Re}[\vec{\vec{\lambda}}_{\vec{z}}]$ and $\text{Im}[\vec{\vec{\lambda}}_{\vec{z}}]$ for time-reversal symmetric systems with $k^2/(\Delta k_n^2 Q) \gg 1$ are related to $k^2/(\Delta k_n^2 Q)$ through,

$$\sigma^2_{\text{Re}[\vec{\vec{\lambda}}_{\vec{z}}]} = \sigma^2_{\text{Im}[\vec{\vec{\lambda}}_{\vec{z}}]} = \frac{1}{\pi}\frac{1}{k^2/(\Delta k_n^2 Q)} \quad (5)$$

This relation has been verified experimentally in Ref. **[23]** for a one-port cavity and will be assumed to hold true for the 2-port results discussed in this paper for data-sets with $k^2/(\Delta k_n^2 Q) > 5$. For data-sets with $k^2/(\Delta k_n^2 Q) < 5$, the following procedure is employed. First, we numerically generate marginal PDFs of the real and imaginary parts of the normalized impedance eigenvalues using random-matrix Monte-Carlo simulations with square matrices of size $N = 1000$, and the value of $k^2/(\Delta k_n^2 Q)$ in the simulations ranging from 0.1 to 5 in steps of 0.1. We determine the variance ($\sigma^2$) of these numerically generated PDFs and fit their dependence on $k^2/(\Delta k_n^2 Q)$ to a polynomial function $\sigma^2 = \Theta(k^2/(\Delta k_n^2 Q))$ of high order. We then determine the variance of the PDF of the real part, i.e. $\sigma^2_{\text{Re}[\vec{\vec{\lambda}}_{\vec{z}}]}$ (which is equal to the variance of the PDF of the imaginary part $\sigma^2_{\text{Im}[\vec{\vec{\lambda}}_{\vec{z}}]}$ to good approximation **[20, 21, 23]**) of the experimentally-determined normalized impedance eigenvalues and solve the inverse polynomial function $k^2/(\Delta k_n^2 Q) = \Theta^{-1}(\sigma^2_{\text{Re}[\vec{\lambda}_{\vec{z}}]})$ to obtain a unique estimate of $k^2/(\Delta k_n^2 Q)$ corresponding to that experimental data-set.

This paper is organized into the following sections. Section II explains the experimental setup and the data acquisition process. By using the measured frequency-dependent radiation impedance matrix, we carry out the normalization of the cavity impedance to uncover the universally fluctuating $\vec{\vec{z}}$; which in turn yields $\vec{\vec{s}}$ and the normalized admittance matrix $\vec{\vec{y}}$. Section III is divided into four sub-sections and presents our experimental results on the universal fluctuations in the eigenvalues of $\vec{\vec{s}}$, $\vec{\vec{z}}$ and $\vec{\vec{y}}$. Firstly, in sub-section III A, the statistical independence of the magnitude and phase of the eigenvalues of $\vec{\vec{s}}$ is experimentally established. The marginal distributions for the magnitude and phase of the eigenvalues of $\vec{\vec{s}}$ are then compared with predictions from RMT. Sub-section III B then explores the evolution of the joint PDF of the $\vec{\vec{s}}$-eigenphases as a function of increasing loss. In sub-section III C, we experimentally test the predictions for the joint PDF of $\vec{\vec{s}}\vec{\vec{s}}^{\dagger}$ (where † denotes the conjugate transpose) from Ref. **[31]** as a function of cavity loss. Sub-section III D then shows experimental data testing the similarity in the PDFs for the eigenvalues of $\vec{\vec{z}}$ and $\vec{\vec{y}}$ and also compares these experimentally obtained PDFs with those from RMT. A technical issue encountered in these 2-port experiments is the presence of non-zero, off-diagonal terms in the measured $\vec{\vec{Z}}_{rad}$. These terms account for the direct-path processes ("cross-talk") between the two ports and come about because of the finite physical separation between the two-ports in the experiment during the radiation measurement. The role of these non-zero, off-diagonal $\vec{\vec{Z}}_{rad}$ terms in determining the universal PDFs of $\vec{\vec{z}}$ is explained in Section IV. Section V concludes this paper with a summary of our experimental findings.

## II: Experimental Setup, Data Acquisition and Construction of normalized $\vec{\vec{z}}$, $\vec{\vec{s}}$ and $\vec{\vec{y}}$:

Microwave-cavity resonators with irregular shapes (where the classical ray trajectories are chaotic) have proved to be a favored test-bed to validate statistical predictions on chaotic scattering **[7]**. In this paper, we present findings on an air-filled, quarter bow-tie shaped billiard cavity [Fig.1 (a)] driven by two-ports. The cavity is 7.87 mm deep and behaves as a two-dimensional resonator when the driving frequency is less than 19.05 GHz. The curved walls ensure that the ray trajectories are chaotic and that there are only isolated classically periodic-orbits **[32]**. Experimental studies on the eigenvalue statistics **[33]**, eigenfunction statistics **[34, 35]**, 1-port impedance **[23]** and 1-port scattering statistics **[24]** as well as an impedance-based Hauser-Feschbach-type relation **[36]** have produced good agreement with theoretical predictions based on RMT.

To set up the investigation, we introduce two driving ports [Fig.1(b)] which are placed roughly 20 cm apart, and are labeled Port-1 and Port-2. The ports are located sufficiently far away from the side-walls of the cavity so that the near-field structure of each port is not altered by the walls. Both ports are sections of coaxial transmission lines, where the exposed center-conductor extends from the top plate of the cavity and makes contact with the bottom plate, injecting current into the bottom plate [Fig.1(c)]. The ports are non-identical; the diameter of the inner conductor is 2a=1.27 mm for Port-1 and 2a=0.635 mm for Port-2.

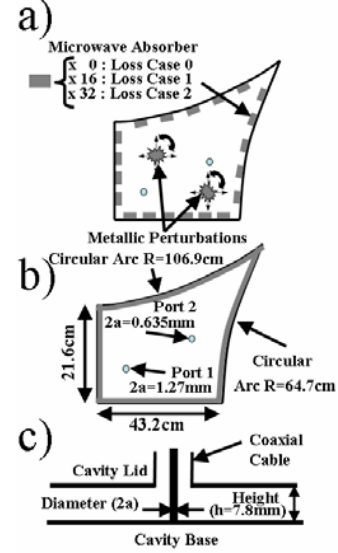

**Fig.1: (a) Top view of quarter-bow-tie microwave cavity used for the experimental "Cavity Case". The two perturbations with serrated edges are shown as the gray shapes. The small, gray, uniformly-spaced rectangles lining the side walls of the cavity represent 2cm-long strips of microwave absorber which are used to control the loss in the cavity.(Loss Case 0 : 0 strips, Loss Case 1: 16 strips, Loss Case 2: 32 strips). (b) The implementation of the experimental "Radiation Case" is shown. The gray lining on the side walls is a homogenous layer of microwave absorber ~ 2 mm thick. The physical dimensions of the cavity are shown in the schematic. The approximate locations of the two driving-ports are also shown. (c) Cross-section view of both driving-ports inside the cavity. The cavity is 7.87 mm in depth. The diameter of the inner conductor is 2a (=1.27 mm for Port 1; =0.635 mm for Port 2).**

As in our previous studies **[23, 24]**, the normalization of the measured data is a two-step procedure. The first step, what we refer to as the "Cavity Case" involves measuring a large ensemble of the full-2x2 scattering matrix $\vec{\vec{S}}_{cav} = \begin{bmatrix} S_{11} & S_{12} \\ S_{21} & S_{22} \end{bmatrix}$ using an Agilent E8364B Vector Network Analyzer. To realize this large ensemble, two metallic perturbers (shown as gray solids in Fig.1.(a) ), each of typical dimensions 6.5 cm x 4 cm x 0.78 cm are used. The perturbers are roughly the order of a wavelength in size at 5 GHz. The edges of the perturbers are intentionally serrated to further randomize the wave scattering within the cavity by preventing the formation of standing waves between the straight wall segments of the cavity and the edges of the perturbations. The perturbers are systematically translated and rotated through one hundred different locations within the volume of the cavity. Hence each orientation of the two perturbers results in a different internal field structure within the cavity. Thus we measure one hundred cavity configurations all having the same volume, coupling geometry for the driving ports, and almost exactly the same cavity conduction loss. For each configuration of the perturbers, $\vec{\vec{S}}$ is measured as a function of frequency from 3 to 18 GHz in 16000 equally spaced steps. Thus, an ensemble of 1,600,000 cavity scattering matrices $\vec{\vec{S}}$ is collected. Special care is taken not to bring the perturbers too close to the ports so as not to alter the near-field structure of the ports.

The dominant loss mechanism in the empty cavity is due to ohmic losses in the broad top and bottom plates of the cavity. The fluctuations in loss from mode-to-mode are small and come from differences in field configurations around the side walls. The degree of loss can be increased in a controlled manner by partially lining the inner side-walls with 2 cm-long strips of microwave absorber [Fig.1(a)] having uniform spacing. Three lossy Cavity Cases – labeled "Loss Case 0" : with no absorbing strips, "Loss Case 1" : with 16 absorbing strips and "Loss Case 2" : with 32 absorbing strips, are measured. Along with frequency, the three "loss cases" lead to an experimental control over the value of $k^2/(\Delta k_n^2 Q)$ from 0.9 to 28.

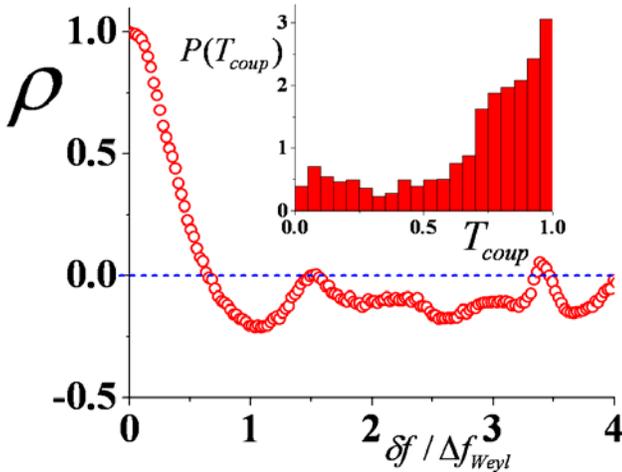

Fig.2: (Color Online) Spectral correlation function
$$\rho(\delta f) = \frac{<|S_{11}(f_o)||S_{11}(f_o+\delta f)|> - <|S_{11}(f_o)|><|S_{11}(f_o+\delta f)|>}{\sigma_{|S_{11}(f_o)|}\sigma_{|S_{11}(f_o+\delta f)|}}$$
of the measured cavity reflection coefficient. Each red-circle symbol represents the correlation between the one hundred different renditions of the Loss Case 0 cavity |S$_{11}$| at frequency $f_o = 3GHz$ with the one hundred different renditions of the same cavity |S$_{11}$| at frequency $f_o + \delta f$. The mean mode-spacing is determined to be $\Delta f_{Weyl} \approx 42 MHz$. Inset: The PDF of the raw-data transmission coefficient of the two ports ($P(T_{coup})$) is shown for Loss Case 0 cavity from 3 to 18 GHz. Note the broad range of coupling values present in the un-normalized data.

To make a quantitative assessment of the degree of "non-ideal coupling" (mismatching) of the two ports with the cavity, we compute the transmission coefficient $T_{coup}$ of the ports [16] as a function of frequency from 3 to 18 GHz. We define $T_{coup} = 1 - |\hat{\lambda}_{<\vec{\vec{S}}>}|^2$, where $\hat{\lambda}_{<\vec{\vec{S}}>}$ are the two complex scalar eigenvalues of $<\vec{\vec{S}}>$. Here, $<\vec{\vec{S}}>$ is the average over the measured ensemble of $\vec{\vec{S}}$ at each frequency. $T_{coup} = 1(0)$ represents the case when the ports are perfectly matched (mismatched) to the cavity. The inset in Fig.2 shows the PDF of the measured $T_{coup}$ (i.e. $P(T_{coup})$) for a Loss Case 0 cavity from 3 to 18 GHz. The PDF is fairly widely spread over the range 0 to 1 with a mean value of ~0.7, and with a standard deviation of ~0.3. An analysis of the coupling and loss for the scattering matrix in similar microwave cavities is presented in Ref. [37].

The degree to which the two perturbations produce a change in the internal field structure of the cavity can be qualitatively inferred by looking at the frequency correlations in the measured $\vec{\vec{S}}$ data. In Fig. 2 for Loss Case 0, the frequency correlation function
$$\rho(\delta f) = \frac{<|S_{11}(f_o)||S_{11}(f_o+\delta f)|> - <|S_{11}(f_o)|><|S_{11}(f_o+\delta f)|>}{\sigma_{|S_{11}(f_o)|}\sigma_{|S_{11}(f_o+\delta f)|}}$$
with $f_o = 3GHz$ is shown as the red circles. The averaging $<...>$ is done over the one hundred different configurations of the perturbations inside the cavity, and $\sigma_{|S_{11}(f)|}$ represents the standard deviation of the one hundred different measurements for cavity $|S_{11}(f)|$ at frequency $f$. The frequency $f_o = 3GHz$ represents the lowest of the frequencies that we experimentally tested; and therefore the worst-case scenario for performing the approximation to true ensemble averaging. Based on the area and perimeter of the cavity, the Weyl formula [30] yields a typical mean-spacing of $\Delta f_{Weyl} \cong 42MHz$ between the eigenmodes of the cavity around $f_o$. From Fig.2, it is observed that the experimentally determined correlations in frequency die off within one mean-spacing $\Delta f_{Weyl}$. However, the correlation function in Fig.2 is similar to those obtained under local, rather than global, perturbations of the system [38]. We have previously identified the fact that short ray orbits inside the cavity will produce systematic deviations of the finite ensemble average from a true ensemble average [20]. We therefore invoke ergodicity and also employ frequency averaging of the data. Since the frequency averaging ranges that we use below are very much larger than $\Delta f_{Weyl}$ (typically by a factor of ~20), this confirms that our frequency (in addition to perturber configuration) averaging is an effective means of approximating a true ensemble average.

The second step of our normalization procedure is what we refer to as the "Radiation Case" [Fig.1(b)]. In this step, the side-walls of the cavity are completely lined with ~ 2 mm thick microwave absorber (ARC-Tech DD 10017) which gives about 20-25 dB reflection loss between 3 and 18 GHz for normal incidence. The perturbers are removed so as not to produce any reflections back to the ports. Port-1 and Port-2 are left untouched- so that they retain the same coupling geometry as in the "Cavity Case". The radiation measurement now involves measuring the resultant 2x2-scattering matrix, which we label $\vec{\vec{S}}_{rad} = \begin{bmatrix} S_{11rad} & S_{12rad} \\ S_{21rad} & S_{22rad} \end{bmatrix}$, from 3 to 18 GHz with the same 16000 frequency steps as in the "Cavity Case". The microwave absorber serves to severely suppress any reflections from the side-walls. This effectively simulates the situation of the side-walls of the cavity being moved out to infinity (radiation-boundary condition). The off-diagonal terms in $\vec{\vec{S}}_{rad}$ correspond to direct-path processes between the two ports. The contribution of these terms has been taken into account in the analysis and results that follow (Section III (A,B,C,D)). The hazards associated with ignoring these terms in the normalization process deserves special mention and are discussed in Section IV.

Having measured the ensemble of cavity $\vec{\vec{S}}$ and the corresponding radiation $\vec{\vec{S}}_{rad}$, we convert these quantities into the corresponding cavity impedance $\vec{\vec{Z}}$ and radiation impedance $\vec{\vec{Z}}_{rad}$ matrices respectively using Eq. (4), where each port has a single operating mode with characteristic impedance of $50\Omega$ over the frequency range of the experiment.

Every measured $\vec{\vec{Z}}$ is then normalized with the corresponding measured $\vec{\vec{Z}}_{rad}$ at the same frequency using Eq. (2). Having obtained the normalized impedance matrix $\vec{\vec{z}}$, it is then converted to the normalized scattering matrix $\vec{\vec{s}}$ using Eq. (3); and the normalized admittance matrix $\vec{\vec{y}}$ ($\vec{\vec{y}} = \vec{\vec{z}}^{-1}$). These normalized quantities represent the corresponding electromagnetic response of the chaotic-cavity in the limit of perfect coupling between the driving ports and the cavity over the entire frequency range of the experiment from 3 to 18 GHz.

### III: Experimental Results:

In this section, we give our experimental results on the universal statistical fluctuations in the eigenvalues of $\vec{\vec{s}}$, $\vec{\vec{z}}$ and $\vec{\vec{y}}$. Each 2x2 $\vec{\vec{s}}$, $\vec{\vec{z}}$ or $\vec{\vec{y}}$ yields two complex eigenvalues – which possess certain universal statistical properties in their marginal and joint PDFs.

### III A: Statistical Independence of $|\lambda_{\vec{s}}|$ and $\phi_{\lambda_{\vec{s}}}$

Having obtained the ensemble of normalized $\vec{\vec{s}}$, we diagonalize $\vec{\vec{s}}$ using an eigenvalue decomposition, $\vec{\vec{s}} = \vec{\vec{V}}_{\vec{s}} \vec{\vec{\lambda}}_{\vec{s}} \vec{\vec{V}}_{\vec{s}}^{-1}$; where, $\vec{\vec{V}}_{\vec{s}}$ is the 2x2 eigenvector matrix of $\vec{\vec{s}}$; and $\vec{\vec{\lambda}}_{\vec{s}}$ is a diagonal matrix containing the two complex eigenvalues of $\vec{\vec{s}}$. In the time-reversal symmetric, lossless limit, $\vec{\vec{s}}$ is unitary. This dictates that $\vec{\vec{V}}_{\vec{s}}$ be an orthogonal matrix and $\vec{\vec{\tilde{\lambda}}}_{\vec{s}} = \begin{bmatrix} Exp[i\tilde{\phi}_1] & 0 \\ 0 & Exp[i\tilde{\phi}_2] \end{bmatrix}$. In the presence of loss, $\vec{\vec{V}}_{\vec{s}}$ is no longer orthogonal and $\vec{\vec{s}}$ now has complex, sub-unitary eigenvalues, i.e. $\vec{\vec{\lambda}}_{\vec{s}} = \begin{bmatrix} |\lambda_1|e^{i\phi_1} & 0 \\ 0 & |\lambda_2|e^{i\phi_2} \end{bmatrix}$, where $|\lambda_{1,2}| < 1$. Reference [24] has shown that for a 1-port system, the magnitudes and phases of the normalized 1-port scattering coefficient $s$ are statistically independent. The independence was shown to be extremely robust and is unaffected by the presence of loss. For a two-port setup, as in the experiments presented in this paper, this would imply statistical independence of the magnitude and phases of the eigenvalues of $\vec{\vec{s}}$.

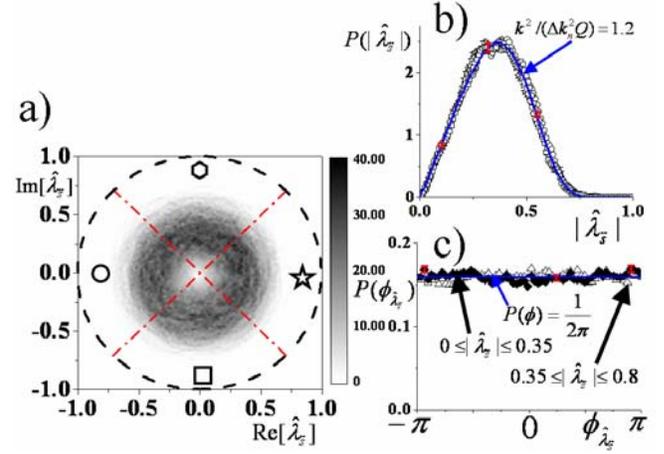

Fig.3: (Color Online) (a) The density of eigenvalues of $\hat{\lambda}_{\vec{s}}$ in the complex plane is shown for frequencies in the range 7.6 GHz to 8.1 GHz for Loss Case 0. The gray-scale code white, light gray, dark gray, black are in ascending density order. (b) Angular slices ($90^o$) with the symbols (stars, hexagons, circles, squares) indicate regions where the PDF of $|\hat{\lambda}_{\vec{s}}|$ of the data in (a) is calculated and shown. Observe that the four PDFs are nearly identical. The blue solid line is the numerical prediction from Random Matrix Theory using the loss parameter $k^2/(\Delta k_n^2 Q) = 1.2$. The red error bars indicate the statistical binning error in the histograms. (c) Experimental histogram approximations to the PDF of the eigenphase of $\vec{\vec{s}}$ (i.e., $\phi_{\hat{\lambda}_{\vec{s}}}$). Two annular rings defined by $0 \leq |\hat{\lambda}_{\vec{s}}| \leq 0.35$ and $0.35 \leq |\hat{\lambda}_{\vec{s}}| \leq 0.8$ of the data in (a) are taken and the histograms of the phase of the points within these regions are shown as the solid diamonds and hollow triangles respectively. The red error bars indicate typical statistical binning errors for the data. The blue solid line is a uniform distribution ($P(\phi) = 1/(2\pi)$).

To test this hypothesis, the two complex eigenvalues of the $\vec{\vec{s}}$ ensemble are grouped into one list, which we shall refer to as "$\hat{\lambda}_{\vec{s}}$". We observe that grouping the two eigenvalues together as opposed to randomly choosing one of the two eigenvalues does not change the statistical properties of the results that follow. Figure 3 (a) shows a plot in the complex plane of the eigenvalue density for a representative set of measured $\vec{\vec{s}}$ ranging between 7.6 to 8.1 GHz where the loss-parameter is roughly constant. The gray-scale level at any point in Fig.3(a) indicates the number of points for $\{ \text{Re}[\hat{\lambda}_{\vec{s}}], \text{Im}[\hat{\lambda}_{\vec{s}}] \}$ that lies within a local rectangular region of size 0.01 x 0.01. Next, angular slices which subtend a polar-angle of $\pi/2$ are taken and histogram approximations to the PDF of $|\hat{\lambda}_{\vec{s}}|$ of the points lying inside each of the four slices are computed. This is shown by the stars, hexagons, circles and squares in Fig. 3(b). It can be observed that the PDF approximations are essentially identical and independent of the angular-slice. By grouping the real part of the eigenvalues of $\vec{z}$ in to one list and computing its variance (i.e. $\sigma^2_{\text{Re}[\hat{\lambda}_{\vec{z}}]}$), we solve the inverse polynomial function $k^2/(\Delta k_n^2 Q) = \Theta^{-1}(\sigma^2_{\text{Re}[\hat{\lambda}_{\vec{z}}]})$ to yield an estimate of $k^2/(\Delta k_n^2 Q) = 1.2 \pm 0.1$ for this data set. The blue solid line shows the numerical RMT prediction [39] which is computed using a single value of $k^2/(\Delta k_n^2 Q) = 1.2$. The red error bars in Fig. 3(b) which are representative of the typical statistical binning error of the experimental histograms show that the data agrees well with the numerical RMT PDF.

In Fig.3(c), the histogram approximations of the phase of the points lying within two-annular rings defined by $0 \leq |\hat{\lambda}_{\vec{s}}| \leq 0.35$ (solid diamonds) and $0.35 \leq |\hat{\lambda}_{\vec{s}}| \leq 0.8$ (hollow triangles) are shown. A nearly uniform distribution is obtained for both cases indicating that the PDF of the phase of $\hat{\lambda}_{\vec{s}}$ is independent of the radius of the annular ring. Also shown in blue is the uniform distribution with $P(\phi) = 1/(2\pi)$. Figure 3 thus supports the hypothesis that the magnitude and phase of the eigenvalues of $\vec{s}$ are statistically independent of each other and that the eigen-phase of $\vec{s}$ is uniformly distributed from $0$ to $2\pi$.

### III B: Joint PDFs of eigenphases of $\vec{s}$.

Section III A has established the uniform distribution of the marginal PDF of the eigenphases of $\vec{s}$. Here we explore the statistical inter-relationships between the two eigenphases of $\vec{s}$ by looking at their joint PDFs i.e., $P(\phi_1, \phi_2)$. In the lossless limit the eigenvalues of $\vec{s}$ are of unit modulus and their marginal distribution is uniform in phase along the unit-circle. Reference [8] has shown that the joint PDF of the eigenphases $\phi_1$ and $\phi_2$, shows a clear anti-correlation, i.e. $P(\phi_1, \phi_2) \propto |e^{i\phi_1} - e^{i\phi_2}|^\beta$, where $\beta = 1(2)$ for a time-reversal(broken) GOE(GUE) system. In the lossless GOE case this anti-correlation is $<\phi_1 \phi_2> \cong -0.216$, where $-\pi \leq \phi_{1,2} \leq \pi$ [21]. As losses are introduced, the eigenvalues of $\vec{s}$ are no longer confined to move along the unit-circle; but rather are distributed inside the unit circle in a manner dependent upon the loss in the system (as was shown in Fig.3(a)). The sub-unitary modulus of the eigenvalues thus presents an extra degree of freedom for eigenvalue avoidance, hence we expect a reduced anti-correlation of the eigenphases as the losses increase. To our knowledge, there exists no analytic formula for the evolution of the joint PDF of the eigenphases of $\vec{s}$ as a function of loss. In the following paragraphs, we thus compare our experimental results for the joint PDF of the eigenphases of $\vec{s}$ with numerical computations of results from RMT [39].

In order to make comparisons of the data with numerical computations from RMT, we transform the eigenphases $\phi_1$ and $\phi_2$ to $\kappa_1$ and $\kappa_2$, as follows,

$$\kappa_1 = \phi_1 - \phi_2 - \pi + 2\pi H(\phi_2 - \phi_1)$$
$$\kappa_2 = \phi_2 \qquad (6)$$

where $H(x)$ is the Heaviside step function ($H(x) = 0$ for $x < 0$; $H(x) = 1$ for $x > 0$). This transformation of variables has the effect of making $\kappa_1$ and $\kappa_2$ statistically independent, with all the correlation information between $\phi_1$ and $\phi_2$ being contained in $\kappa_1$; and $\kappa_2$ being uniformly distributed (as shown in Fig.3(c)). In the lossless case, it can be easily deduced from $P(\phi_1, \phi_2) \propto |e^{i\phi_1} - e^{i\phi_2}|^\beta$, that $P(\kappa_1) = Cos(\kappa_1/2)/4$ for $\beta = 1$.

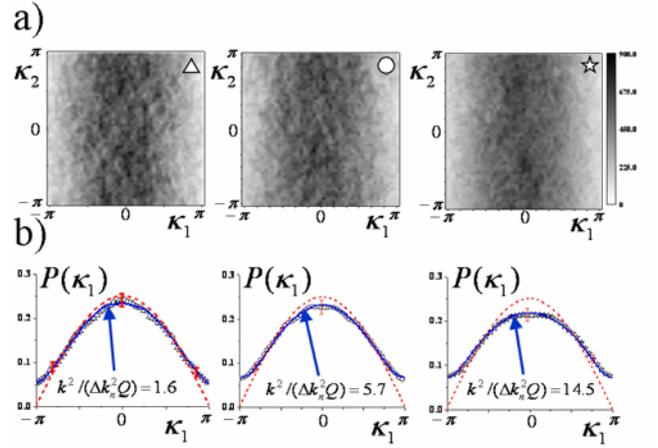

Fig.4: (Color Online) (a) The joint PDF ($P(\kappa_1, \kappa_2)$) of the transformed eigenphases $\kappa_1$ and $\kappa_2$ for Loss Case 0 (triangles: left), Loss Case 1 (circles: center) and Loss Case 2 (stars: right) in the frequency range of 10.4-11.7 GHz. The gray-scale code white, light gray, dark gray, black are in ascending density order. (b) Marginal PDFs for $\kappa_1$ (Loss Case 0 (triangles: left), Loss Case 1 (circles: center) and Loss Case 2 (stars: right)) of the data shown in the top row. The dashed red line is the lossless prediction $P(\kappa_1) = Cos(\kappa_1/2)/4$. The blue solid lines are the numerical RMT prediction for $P(\kappa_1)$ with $k^2/(\Delta k_n^2 Q)$ =1.6 (left); 5.7 (center) and 14.5 (right).

The top row of Fig.4 shows the density plots of $\kappa_1$ and $\kappa_2$ for the three different loss-cases (Loss-Case 0: triangles, Loss-Case 1: circles, Loss-Case 2: stars) in the frequency range of 10.4-11.7 GHz. This corresponds to $k^2/(\Delta k_n^2 Q) = 1.6 \pm 0.1, 5.7 \pm 0.1$ and $14.5 \pm 0.1$ respectively. For the data set represented by the triangles, the value of $k^2/(\Delta k_n^2 Q)$ was determined by computing the variance of the real part of the grouped eigenvalues of $\vec{z}$ (i.e. $\sigma^2_{\text{Re}[\hat{\lambda}_{\vec{z}}]}$) and solving the inverse polynomial function $k^2/(\Delta k_n^2 Q) = \Theta^{-1}(\sigma^2_{\text{Re}[\hat{\lambda}_{\vec{z}}]})$. For the data sets represented by the circles and stars, the value of $k^2/(\Delta k_n^2 Q)$ was determined by computing the variance of the real part of the grouped eigenvalues of $\vec{\vec{z}}$ and Eq.(5). As the plots indicate, the statistical variation is entirely contained in the $\kappa_1$ direction, with $\kappa_2$ being nearly uniformly distributed. The gray-scale on the plots indicates the number of points for $\{\kappa_1, \kappa_2\}$ which lie within a local rectangular region of size 0.01 x 0.01. The corresponding anti-correlation of the eigenphases $<\phi_1, \phi_2> \cong -0.17, -0.16, -0.15 \, (-\pi \leq \phi_{1,2} \leq \pi)$ for the triangles, circles and stars respectively.

The bottom row of Fig. 4 shows histogram approximations to the Marginal PDFs of $\kappa_1$ for all three cases of loss (Loss-Case 0: triangles, Loss-Case 1: circles, Loss-Case 2: stars) for the data shown in the top row. The blue solid line is the numerical RMT computation for $P(\kappa_1)$ which is based upon the loss parameters stated above. The red dashed-line is the predicted PDF of $\kappa_1$ in the lossless case. The red error-bars indicate the typical statistical binning error for the experimental PDF histograms. The agreement between the experimentally determined $P(\kappa_1)$ (symbols) and the numerically generated $P(\kappa_1)$ (blue trace) is good and well within the error-estimates. We observe that as the losses increase, the histograms for $P(\kappa_1)$ tends to grow progressively wider and develop smooth tails - which results in a reduced anti-correlation between $\phi_1$ and $\phi_2$, as expected.

### III C: Joint PDF of eigenvalues of $\vec{\vec{s}}\vec{\vec{s}}^{\dagger}$

We now consider the joint PDF of the eigenvalues of $\vec{\vec{s}}\vec{\vec{s}}^{\dagger}$, where † denotes the conjugate transpose. Since $\vec{\vec{s}}\vec{\vec{s}}^{\dagger}$ is Hermitian, its eigenvalues are purely real. The matrix $\vec{\vec{s}}\vec{\vec{s}}^{\dagger}$ is of significant interest in the quantum-transport community as it determines the conductance fluctuations of ballistic quantum-dots in the presence of dephasing/loss. Owing to the analogy between the time-independent Schrödinger equation and the two-dimensional Helmholtz equation, the microwave billiard experiment presents itself as an ideal platform to test statistical theories for these quantum fluctuations without the complicating effects of thermal smearing [40] and Coulomb interactions, as discussed in Ref. [41].

Models have been introduced to quantify the loss of quantum phase coherence (dephasing) of transport electrons in quantum dots [42, 43, 44, 45]. These models generally utilize a fictitious lead attached to the dot that has a number of channels $N_\phi$ each with transparency $\Gamma_\phi$. Electrons that enter one of the channels of this lead are re-injected into the dot with a phase that is uncorrelated with their initial phase, and there is no net current through the fictitious lead. An alternative model of electron transport employs a uniform imaginary term in the electron potential [46, 47], leading to loss of probability density with time, similar to the loss of microwave energy in a cavity due to uniformly distributed losses in the walls and lids. As far as the conductance is concerned, it was shown that these two models are equivalent in the limit when the number of channels in the dephasing lead $N_\phi \to \infty$ and $\Gamma_\phi \to 0$, with the product $\gamma = N_\phi \Gamma_\phi$ remaining finite [31, 43, 48]. In this case, the dephasing parameter $\gamma$ is equivalent to a loss parameter describing the strength of uniformly distributed losses in the system. Other models have been proposed that consider parasitic channels [43, 49] or an "absorbing patch" or "absorbing mirror" [50] to describe losses in a microwave cavity. Here we examine the predictions of Brouwer and Beenakker using the dephasing lead model in the limit mentioned above. In this case the dephasing parameter $\gamma$ is treated as a loss parameter describing fairly uniformly distributed losses in our microwave cavity, and is found to be proportional to the loss parameter $k^2/(\Delta k_n^2 Q)$ that we introduced in this and other publications [23, 24, 41, 51].

Reference [31] has shown that the eigenvalues of $\vec{\vec{s}}\vec{\vec{s}}^{\dagger}$ can be denoted as $1-T_1$ and $1-T_2$ (where $T_1$ and $T_2$ determine the absorption strength of this fictitious port) with the statistical properties of $T_1$ and $T_2$ dependent on the parameter $\gamma$. When $\gamma = 0$, $T_1$ and $T_2$ equal zero and $\vec{\vec{s}}$ is unitary. As $\gamma$ increases, $T_1$ and $T_2$ migrate towards 1. Equation 17(a) (Eq.(7) below) and Eq.17(b) of Ref. [31] are exact analytic expressions for the joint PDF of $T_1$ and $T_2$ in terms of $\gamma$ for both the GOE and GUE cases respectively. At all values of $\gamma$, the analytic expression for $P(T_1, T_2; \gamma)$ shows strong anti-correlation between $T_1$ and $T_2$ [31],

$$P(T_1, T_2; \gamma) =$$
$$\frac{1}{8} T_1^{-4} T_2^{-4} Exp(-\frac{1}{2}\gamma(T_1^{-1} + T_2^{-1})) |T_1 - T_2| (\gamma^2(2 - 2e^\gamma + \gamma + \gamma e^\gamma)$$
$$- \gamma(T_1 + T_2)(6 - 6e^\gamma + 4\gamma + 2\gamma e^\gamma + \gamma^2)$$
$$+ T_1 T_2 (24 - 24e^\gamma + 18\gamma + 6\gamma e^\gamma + 6\gamma^2 + \gamma^3)).$$

(7)

For our experiment, once the ensemble of $\vec{\vec{s}}$ has been obtained, $T_1$ and $T_2$ can be easily determined by computing the eigenvalues of $\vec{\vec{s}}\vec{\vec{s}}^{\dagger}$. In Fig. 5, contour density plots of $P(T_1, T_2)$ is shown for the Loss Case 0 (Fig. 5(a) : 3.2-4.2 GHz) and Loss Case 0 (Fig. 5(b): 13.5-14.5 GHz). This corresponds to $k^2/(\Delta k_n^2 Q)$ values of $1 \pm 0.1$ for Fig. 5(a) and $2.9 \pm 0.1$ for Fig. 5(b). These values of $k^2/(\Delta k_n^2 Q)$ are determined from estimating the variance of the real part of the grouped eigenvalues of $\vec{\vec{z}}$ (i.e. $\sigma^2_{\text{Re}[\hat{\lambda}_{\hat{z}}]}$) and solving the inverse polynomial function $k^2/(\Delta k_n^2 Q) = \Theta^{-1}(\sigma^2_{\text{Re}[\hat{\lambda}_{\hat{z}}]})$ for both data sets. The color-scale level indicates the number of points that lie in a local rectangular region of size 0.01 x 0.01 for Fig. 5(a) and 0.005 x 0.005 for Fig.5(b) (note the change in scales for the plots). We observe that as losses increase the cluster of $T_1$ and $T_2$ values which are centered around ~0.75 for Fig. 5(a) migrates towards values of $T_1$ and $T_2$ approaching 1 (Fig. 5(b)). We also observe a strong anti-correlation in $P(T_1,T_2)$ for $T_1 = T_2$. This anti-correlation is manifested in all the data measured at varying degrees of loss from $k^2/(\Delta k_n^2 Q)$ =0.9 to 28.

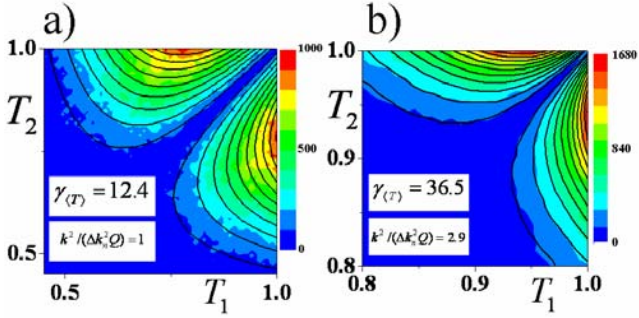

Fig.5: (Color Online) The experimental joint PDF of $T_1$ and $T_2$ (i.e., $P(T_1,T_2)$) for Loss Case 0: 3.2-4.2 GHz (a) and 13.5-14.5 GHz (b). The color codes blue, green, yellow and red are in ascending density order. The black contours are theoretical predictions for $P(T_1,T_2;\gamma)$ obtained from Eq. 7 for $\gamma = 12.4$ (a) and $\gamma = 36.5$ (b).

To estimate the value of $\gamma$ for our experimental data sets, we derive an analytic expression for $\langle T \rangle$ in terms of $\gamma$ from Eq. (7) [41],

$$\langle T \rangle = \frac{1}{4\gamma}(e^{-\gamma}(4(e^\gamma - \gamma - 1) + 4e^\gamma(2e^\gamma - 2 - \gamma(2+\gamma))\xi(-\gamma) \quad (8)$$
$$-2e^{\gamma/2}(e^\gamma(2+\gamma(\gamma-2))-2)\xi(-\gamma/2)))$$

where $\xi(z) = -\int_{-z}^{\infty} \frac{e^{-t}}{t} dt$ is the exponential integral function.

By determining the value of $\langle T \rangle$ from the measured data set, Eq.(8) then uniquely determines the corresponding value of $\gamma \, (\equiv \gamma_{\langle T \rangle})$. This approach yields values of $\gamma_{\langle T \rangle} = 12.4 \pm 0.1$ and $\gamma_{\langle T \rangle} = 36.5 \pm 0.1$ for the data in Fig. 5(a) and Fig. 5(b), respectively. Using these values of $\gamma_{\langle T \rangle}$, we plot the analytic contour curves defined by Eq. (7) for the two loss cases, shown as the solid black lines in Fig. 5. The theoretical curves reflect the same number of contour levels shown in the data. We observe relatively good agreement between the theoretical prediction of Ref. [31] and the experimental data. This agreement between the experimental data and the theoretical prediction is also observed to extend over other loss-cases and frequency ranges. Comparing the value of $k^2/(\Delta k_n^2 Q)$ from each experimental data set with the corresponding value of $\gamma_{\langle T \rangle}$, we empirically determine a linear relation between $k^2/(\Delta k_n^2 Q)$ and $\gamma$, i.e

$\gamma = (12.5 \pm 0.1)k^2/(\Delta k_n^2 Q)$ using 70 points for $\gamma_{\langle T \rangle}$ between ~11 and ~300 [41].

### III D: Marginal PDFs of eigenvalues of $\vec{\vec{z}}$ and $\vec{\vec{y}}$:

In this section we determine the marginal PDFs of the eigenvalues of the normalized impedance $\vec{\vec{z}}$ and normalized admittance $\vec{\vec{y}}$. It has been theorized in [27] that these two quantities have identical distributions for their eigenvalues. References [20, 21] show that attaching an arbitrary lossless two-port network at the interface between the plane of measurement, and the cavity does not alter the statistics of $\vec{\vec{z}}$. If we now assume that this lossless two-port is a transmission line with an electrical-length equal to one-quarter wavelength at the driving frequency, then the lossless two-port acts as an "impedance inverter" [25] thereby presenting a cavity admittance at the plane of measurement. This similarity in the statistical description of $\vec{\vec{z}}$ and $\vec{\vec{y}}$ is predicted to be extremely robust and independent of loss in the cavity, coupling, driving frequency, etc.

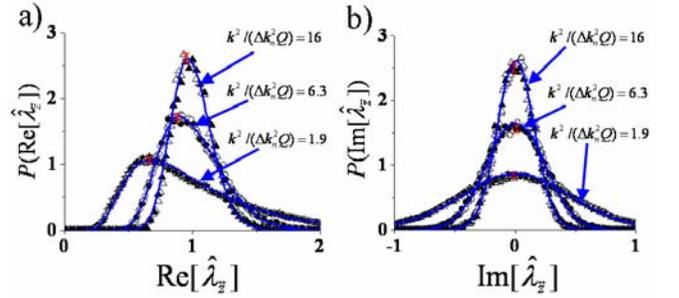

Fig.6: (Color Online) PDFs for the real (a) and imaginary (b) parts of the grouped eigenvalues of the normalized cavity impedance $\hat{\lambda}_{\vec{z}}$ (hollow stars : Loss Case 0; hollow circles : Loss Case 1; hollow triangles : Loss Case 2) in the frequency range of 10.5-12 GHz. The PDFs for the real (a) and imaginary (b) parts of the grouped eigenvalues of the normalized cavity admittance $\hat{\lambda}_{\vec{y}}$ (solid stars : Loss Case 0; solid circles : Loss Case 1; solid triangles : Loss Case 2) in the frequency range of 10.5-12 GHz are also shown. The red error bars indicate the typical statistical binning error of the data. Also shown are the single parameter, simultaneous fits for both impedance and admittance PDFs (blue solid lines), where the loss parameter $k^2/(\Delta k_n^2 Q)$ is obtained from the variance of the data in (a).

For our experimental test of this prediction, we consider the three loss cases, Loss Case 0, 1 and 2, in the frequency range 10.5-12GHz. By an eigenvalue decomposition, each $\vec{\vec{z}}$ and $\vec{\vec{y}}$ matrix yields two complex eigenvalues, which we group together to form $\hat{\lambda}_{\vec{z}}$ and $\hat{\lambda}_{\vec{y}}$ respectively. We observe that grouping the two eigenvalues together as opposed to randomly considering one of the two eigenvalues separately does not alter the statistical results that follow. Histograms of the real and imaginary parts of $\hat{\lambda}_{\vec{z}}$ and $\hat{\lambda}_{\tilde{y}}$ are plotted in Fig. 6. The hollow stars, circles and triangles in Fig.6(a) (Fig.6(b)) correspond to the histogram approximations of the PDF of $\text{Re}[\hat{\lambda}_{\tilde{z}}]$ ($\text{Im}[\hat{\lambda}_{\tilde{z}}]$) for Loss case 0, 1 and 2 respectively. The evolution of these PDFs for $\text{Re}[\hat{\lambda}_{\tilde{z}}]$ and $\text{Im}[\hat{\lambda}_{\tilde{z}}]$ with increasing loss, are in qualitative agreement with the description given in Ref. **[21]**. As losses increase, we observe that the PDFs of $\text{Re}[\hat{\lambda}_{\tilde{z}}]$ shifts from being peaked at $\text{Re}[\hat{\lambda}_{\tilde{z}}] \sim 0.6$ (Loss Case 0) to developing a Gaussian-type distribution that peaks near $\text{Re}[\hat{\lambda}_{\tilde{z}}] \sim 1$ (Loss Case 2). While in Fig. 6(b), as losses increase, the PDFs lose their long tails and become sharper. The solid stars, circles and triangles in Fig. 6(a) (Fig. 6(b)) correspond to the histogram approximations of the PDF of $\text{Re}[\hat{\lambda}_{\tilde{y}}]$ ($\text{Im}[\hat{\lambda}_{\tilde{y}}]$) for Loss case 0, Loss Case 1 and Loss Case 2 respectively. The agreement between the PDF approximations for $\text{Re}[\hat{\lambda}_{\tilde{z}}]$ and $\text{Re}[\hat{\lambda}_{\tilde{y}}]$ ($\text{Im}[\hat{\lambda}_{\tilde{z}}]$ and $\text{Im}[\hat{\lambda}_{\tilde{y}}]$) is good for all the three loss cases. The red error bars are representative of the statistical error introduced from the binning of the data in the histograms. By computing the variance of the PDFs for $\text{Re}[\hat{\lambda}_{\tilde{z}}]$ and by using the inverse polynomial function $k^2/(\Delta k_n^2 Q) = \Theta^{-1}(\sigma^2_{\text{Re}[\hat{\lambda}_{\tilde{z}}]})$, we obtain a loss parameter of $k^2/(\Delta k_n^2 Q) = 1.9 \pm 0.1$ (Loss Case 0-stars); From the variance of the PDFs for $\text{Re}[\hat{\lambda}_{\tilde{z}}]$ and by Eq. (5), we obtain a loss parameter of $k^2/(\Delta k_n^2 Q) = 6.3 \pm 0.1$ (Loss Case 1-circles) and $k^2/(\Delta k_n^2 Q) = 16 \pm 0.1$ (Loss Case 2-triangles). Using these loss parameter values, a Monte Carlo RMT computation **[39]** yields the solid blue lines which simultaneously fit the data shown in both Fig. 6(a) and Fig. 6(b) for the three loss cases. The agreement between the experimentally observed values and the RMT result are in good agreement for all three cases and within the bounds of the error bars.

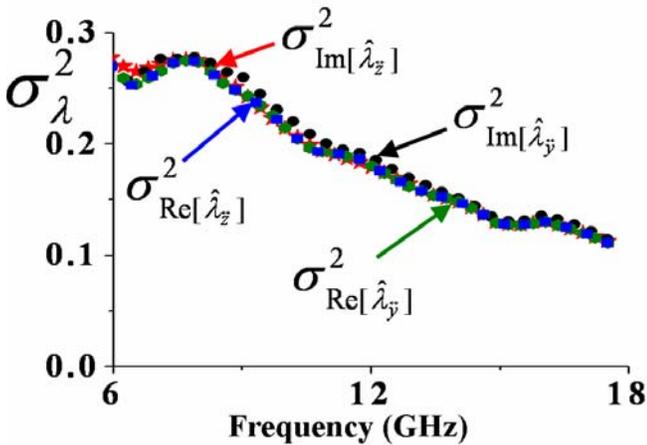

**Fig.7: (Color Online) The variance of $\text{Re}[\hat{\lambda}_{\tilde{z}}]$ (blue squares), $\text{Re}[\hat{\lambda}_{\tilde{y}}]$ (green hexagons); $\text{Im}[\hat{\lambda}_{\tilde{z}}]$ (red stars) and $\text{Im}[\hat{\lambda}_{\tilde{y}}]$ (black circles) distributions are shown as a function of frequency from 6 to 18 GHz for Loss Case 0. The agreement between these four quantities is good and robust over the entire frequency range despite the change in cavity Q.**

We observe that there is a robust agreement between the distributions for $\text{Re}[\hat{\lambda}_{\tilde{z}}]$ and $\text{Re}[\hat{\lambda}_{\tilde{y}}]$ as well as between $\text{Im}[\hat{\lambda}_{\tilde{z}}]$ and $\text{Im}[\hat{\lambda}_{\tilde{y}}]$ over a broad range of frequencies, coupling conditions and loss. To highlight this robust nature, in Fig. 7, we plot the variance of $\text{Re}[\hat{\lambda}_{\tilde{z}}]$ (blue squares), $\text{Re}[\hat{\lambda}_{\tilde{y}}]$ (green hexagons), $\text{Im}[\hat{\lambda}_{\tilde{z}}]$ (red stars) and $\text{Im}[\hat{\lambda}_{\tilde{y}}]$ (black circles) for a Loss Case 0 cavity measurement. Each symbol corresponds to a 1 GHz wide sliding window that steps every 500 MHz over the frequency range from 6 to 18 GHz. It can be seen that the four symbols closely overlap each other over the entire frequency range. The agreement between the symbols (as predicted by Ref. **[20, 21, 27]**) is remarkable despite the variation in coupling, frequency and loss (which varies from $k^2/(\Delta k_n^2 Q) \sim 1$ to 3.5 over this frequency range) within the cavity.

## IV: Importance of the Off-Diagonal Radiation Elements in $\vec{\vec{Z}}_{rad}$

The "radiation impedance" approach to filter out the direct processes involved in a chaotic scattering experiment relies on the accuracy of the measured radiation impedance matrix. This section explains a key technical issue faced while experimentally measuring the radiation impedance matrix of the driving ports; specifically, the presence of non-zero, off-diagonal terms in the measured radiation impedance matrix.

The conjecture that the statistical properties of real-world, physically realizable, wave-chaotic scattering systems can be modeled by an ensemble of large matrices with random elements (governed by certain system symmetries) is applicable only in the semi-classical or short wavelength limit. For the purpose of this conjecture, in the presence of ports, a consistent definition of the short wavelength limit is that, when taking this limit, the size of the ports connecting to the cavity remain constant in units of wavelength. With this definition of the limit, the ratio of the distance between the ports to their size approaches infinity. Thus $\vec{\vec{Z}}_{rad}$ becomes diagonal and approaches a constant at short wavelength.

The conjecture that RMT describes the scattering properties in a specific case assumes that, in the short wavelength limit, rays entering the cavity bounce many times before leaving (i.e., they experience the chaotic dynamics). With the above definition of the short wavelength limit of the ports, this would be the case since the fraction of power reflected back to a port via short (e.g., one or two bounce) paths approaches zero. At finite wavelength, however, it can be anticipated that there could be noticeable deviations from the RMT predictions and that these would be associated with short ray paths. In our experimental determinations of $\vec{\vec{Z}}_{rad}$ we have effectively eliminated the largest source of such non-universal behavior, namely, the short ray paths that go directly between ports 1 and 2. This is the case because these ray paths are already included in our experimental $\vec{\vec{Z}}_{rad}$.

In particular, lining the inner walls of the cavity with microwave absorber for the "Radiation Case" of the experiment, serves to essentially eliminate reflections off the side-walls, but plays no role in suppressing the direct-path interaction (cross-talk) between the two ports. This cross-talk is manifested primarily as non-zero, off-diagonal terms in the measured $\vec{\vec{Z}}_{rad}$ with enhanced frequency dependence relative to the one-port case.

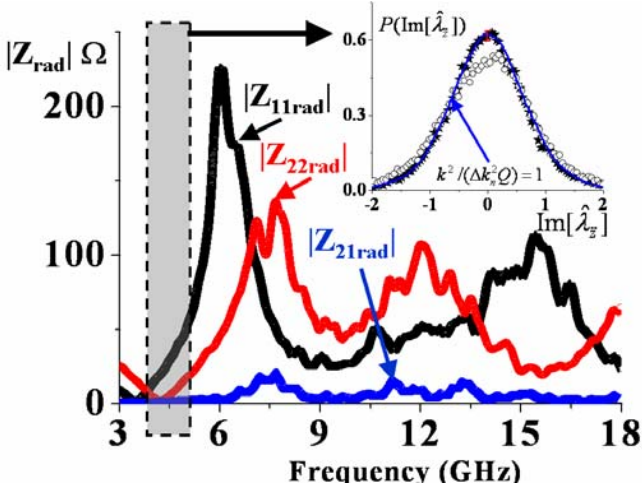

**Fig.8:** (Color Online) Magnitude of the elements of the measured radiation impedance matrix are shown as a function of frequency from 3 to 18 GHz for the setup in Fig.1 (b,c). Inset: PDF of $\text{Im}[\hat{\lambda}_{\tilde{z}}]$ for the Loss Case 0 cavity in the frequency range 4-5 GHz, which is obtained by considering the full 2x2 radiation impedance matrix (stars) and by considering only the contribution of the diagonal elements of the radiation impedance matrix (circles). The blue solid line is the RMT numerical prediction for $k^2/(\Delta k_n^2 Q)$ =1 which is obtained from the variance of the data represented by the stars.

Figure 8 shows the magnitudes of the elements of the radiation impedance matrix $\vec{\vec{Z}}_{rad}$ for the two-port setup shown in Fig. 1(b). Frequency ranges where there is significant cross-talk between the two ports are manifested as large values of $|Z_{21rad}|=|Z_{12rad}|$. Note the complicated structure of the measured elements of $\vec{\vec{Z}}_{rad}$.

To highlight the contribution of short ray paths, the inset of Fig. 8, shows the PDF of the eigenvalues of the normalized impedance for two scenarios of the Loss Case 0 cavity in the 4-5 GHz frequency range. The circles represent the PDF of $\text{Im}[\hat{\lambda}_{\tilde{z}}]$ that is obtained by setting the off-diagonal terms of the measured radiation impedance matrix to zero. The solid stars however, represent the PDF of $\text{Im}[\hat{\lambda}_{\tilde{z}}]$ which is obtained by considering <u>all</u> the elements of the measured radiation impedance matrix during the normalization process (Eq. (2)) to obtain $\vec{\hat{z}}$. The red error bars are representative of the statistical error introduced from the binning of the data in the histograms indicated by the solid stars. We observe a clear discrepancy between the two curves and also note that the PDF represented by the circles does not peak at 0. Using the variance of the measured $\text{Im}[\hat{\lambda}_{\tilde{z}}]$ (stars) and the inverse polynomial function $k^2/(\Delta k_n^2 Q) = \Theta^{-1}(\sigma^2_{\text{Im}[\hat{\lambda}_{\tilde{z}}]})$, we obtain a loss parameter value of $k^2/(\Delta k_n^2 Q) = 1 \pm 0.1$ for this frequency range. We use this value to generate the PDF of $\text{Im}[\hat{\lambda}_{\tilde{z}}]$ using Random Matrix Monte Carlo simulation **[39]**. The resultant numerical prediction is shown as the solid blue line. We observe good agreement between the numerical RMT prediction and the experimentally determined PDF of $\text{Im}[\hat{\lambda}_{\tilde{z}}]$ by considering the full 2x2 radiation impedance matrix. Our choice of the 4-5 GHz range is motivated by the fact that in this range, the ratio of $|Z_{21rad}|/|Z_{22rad}|$ is the largest. This result establishes the importance of off-diagonal terms in $\vec{\vec{Z}}_{rad}$, and helps to validate our approach to removing short-path direct processes between the ports.

### V Summary:

The results discussed in this paper are meant to provide conclusive experimental evidence in support of the "radiation impedance" normalization process introduced in Ref. **[21]** for multiple-port, wave-chaotic cavities. The close agreement between the experimentally determined PDFs and those generated numerically from RMT, support the use of RMT to model statistical aspects of real-world, semi-classical wave-chaotic systems. This paper is a natural two-port extension of the one-port experimental results of Ref. **[23, 24]**. The extension to two-ports makes these results of much broader appeal to other fields of physics and engineering where wave-transport through complex, disordered media is of interest.

We have shown that the full 2x2 radiation impedance matrix of the two-driving ports can accurately quantify the non-ideal and system-specific coupling details between the cavity and the ports as well as the cross-talk between ports, over any frequency range. Hence, given our experimentally-measured, non-ideally coupled cavity data, this normalization procedure allows us to retrieve the universal statistical fluctuations of wave-chaotic systems which are found only in the limit of perfect coupling. We have experimentally tested the evolution of these universal fluctuations traversing from the regime of intermediate to high loss and for different coupling geometries. We find good agreement between the PDFs obtained experimentally to those generated numerically from RMT. Of particular significance is the joint PDF of the eigenphases of $\vec{\vec{s}}$, and the eigenvalues of $\vec{\vec{s}}\vec{\vec{s}}^{\dagger}$ which lead to the universal conductance fluctuations statistics of quantum-transport systems. Our results are not restricted to microwave-billiard experiments but also apply to other allied fields, such as quantum-optics, acoustics and electromagnetic compatibility.

### Acknowledgements:
We acknowledge useful discussions with R. Prange and S. Fishman, as well as comments from Y. Fyodorov, D.V. Savin, P. Brouwer, P. Pereyra and T. Seligman. One of us (S.M.A) acknowledges the hospitality of the Centro Internacional de Ciencias A.C., Cuernavaca, Mexico. This work is supported by the DoD MURI for the study of microwave effects under AFOSR Grant F496200110374, AFOSR DURIP Grants FA95500410295 and FA95500510240, and by the Israel/U.S.A. Binational Science Foundation.